\documentclass[aps,prl,twocolumn,groupedaddress,a4paper,showpacs]{revtex4-1}
\pdfoutput=1

\usepackage{graphicx}
\usepackage[hidelinks]{hyperref}
\usepackage{color}
\usepackage[T1]{fontenc}
\usepackage[utf8]{inputenc}
\usepackage{amssymb}
\usepackage{amsmath}
\usepackage{amsthm}
\usepackage{bbm}
\usepackage{mathtools}
\usepackage{float}
\usepackage{empheq}
\usepackage{rotating}
\usepackage{makecell}

\def\v{\mathbf{v}}
\def\x{\mathbf{x}}
\def\r{\mathbf{r}}
\def\C{\mathcal{C}}
\def\P{\mathcal{P}}
\def\exprho #1{\left\langle #1\right\rangle_\rho}
\def\tr{\operatorname{tr}}
\def\idty{{\leavevmode\rm 1\mkern -5.4mu I}} 

\def\Rl{{\mathbb R}}

\def\PP{{\mathbb P}}\def\SS{{\mathcal S}}

\def\tr{\mathop{\rm tr}\nolimits}

\def\EE{{\mathcal E}}\def\HH{{\mathcal H}}
\def\RR{\mathcal R}

\def\var#1#2{\Delta^2_{#1}(#2)} 

\begin{document}
\title{State-independent uncertainty relations and entanglement detection in noisy systems}
\author{Ren\'e Schwonnek}
\email[]{rene.schwonnek@itp.uni-hannover.de}
\author{Lars Dammeier}
\email[]{lars.dammeier@itp.uni-hannover.de}
\author{Reinhard F. Werner}
\email[]{reinhard.werner@itp.uni-hannover.de}
\affiliation{Leibniz Universit\"at Hannover -  Institut f\"ur Theoretische Physik}

\date{\today}

\begin{abstract}
  Quantifying quantum mechanical uncertainty is vital for the increasing number of experiments that reach the uncertainty limited regime. We present a method for computing tight variance uncertainty relations, i.e., the optimal state-independent lower bound for the sum of the variances for any set of two or more measurements. 
   The bounds come with a guaranteed error estimate, so results of pre-assigned accuracy can be obtained straightforwardly. Our method also works for POVM measurements. Therefore, it can be used for detecting entanglement in noisy environments, even in cases where conventional spin squeezing criteria fail because of detector noise.
\end{abstract}


\maketitle

\section*{Introduction\label{intro}}

Uncertainty relations quantitatively express a phenomenon which is ubiquitous in quantum mechanics: Given two observables $A$ and $B$, it is usually impossible to prepare a state such that the respective outcome distributions of these observables are both sharp. Of course, for the best known example of this, the position and momentum observables, the relation is in every textbook. It was first established by Kennard \cite{Kennard}, who turned Heisenberg's heuristic ideas \cite{Heisenberg} into a quantitative statement. In particular, it was his idea to consider the variances \footnote{For hermitian operators we define the variance of a state $\rho$ by $\var{\rho}{A}= \protect\langle A^2 \protect\rangle_\rho - \protect\langle A \protect\rangle^2_\rho = \tr{\rho A^2 }-(\tr{\rho A })^2$.} of momentum and position in the state $\rho$ as the mathematical expression of sharpness. Kennard's relation $\var\rho P\,\var\rho Q\geq\hbar^2/4$ is tight, i.e., the constant on the right hand side is the best possible, because it is attained for Gaussian pure states.

The aim of our paper is to provide an efficient method to obtain the best possible bounds for any given pair of measurements $A$, $B$. This is of direct use in the increasing number of experiments that reach the uncertainty-limited regime.
A particular application is the certification of entanglement via steering inequalities \cite{Hofmann,Guehne,Guehne2}. In such applications, even if one does not necessarily need an optimal bound, it is crucial to have a correct one, i.e., a bound valid for {\it all} states. Any algorithm based on  computing the uncertainties ``for sufficiently many states'' will fail to guarantee this correctness. In particular, in high dimensional Hilbert spaces, typical states will not have uncertainties near the boundary, so it is actually hard to explore the set of uncertainty pairs  $(\var\rho A , \var\rho B)$ ``from within''. Our method uses instead an ``outer'' approximation, which has the virtue that in every step it provides a correct bound. The bound is iteratively improved, converging to the optimal one. This feature sets our method apart from several recent works, in which ad hoc methods were used to provide uncertainty bounds.
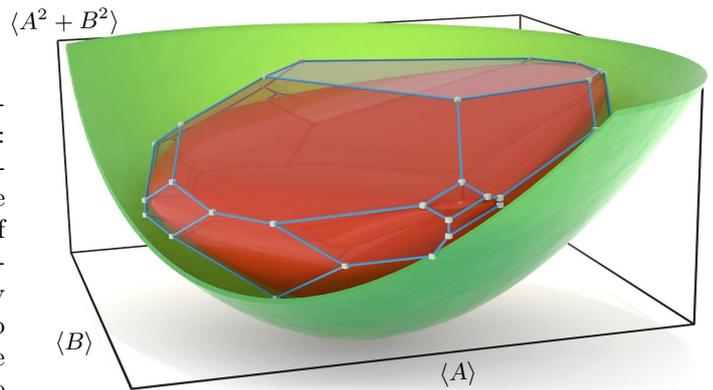
\begin{figure}[t]
\def\svgwidth{\linewidth}
\begingroup%
  \makeatletter%
  \providecommand\color[2][]{%
    \errmessage{(Inkscape) Color is used for the text in Inkscape, but the package 'color.sty' is not loaded}%
    \renewcommand\color[2][]{}%
  }%
  \providecommand\transparent[1]{%
    \errmessage{(Inkscape) Transparency is used (non-zero) for the text in Inkscape, but the package 'transparent.sty' is not loaded}%
    \renewcommand\transparent[1]{}%
  }%
  \providecommand\rotatebox[2]{#2}%
  \ifx\svgwidth\undefined%
    \setlength{\unitlength}{1194.4bp}%
    \ifx\svgscale\undefined%
      \relax%
    \else%
      \setlength{\unitlength}{\unitlength * \real{\svgscale}}%
    \fi%
  \else%
    \setlength{\unitlength}{\svgwidth}%
  \fi%
  \global\let\svgwidth\undefined%
  \global\let\svgscale\undefined%
  \makeatother%
  \begin{picture}(1,0.60281313)%
    \put(0,0){\includegraphics[width=\unitlength]{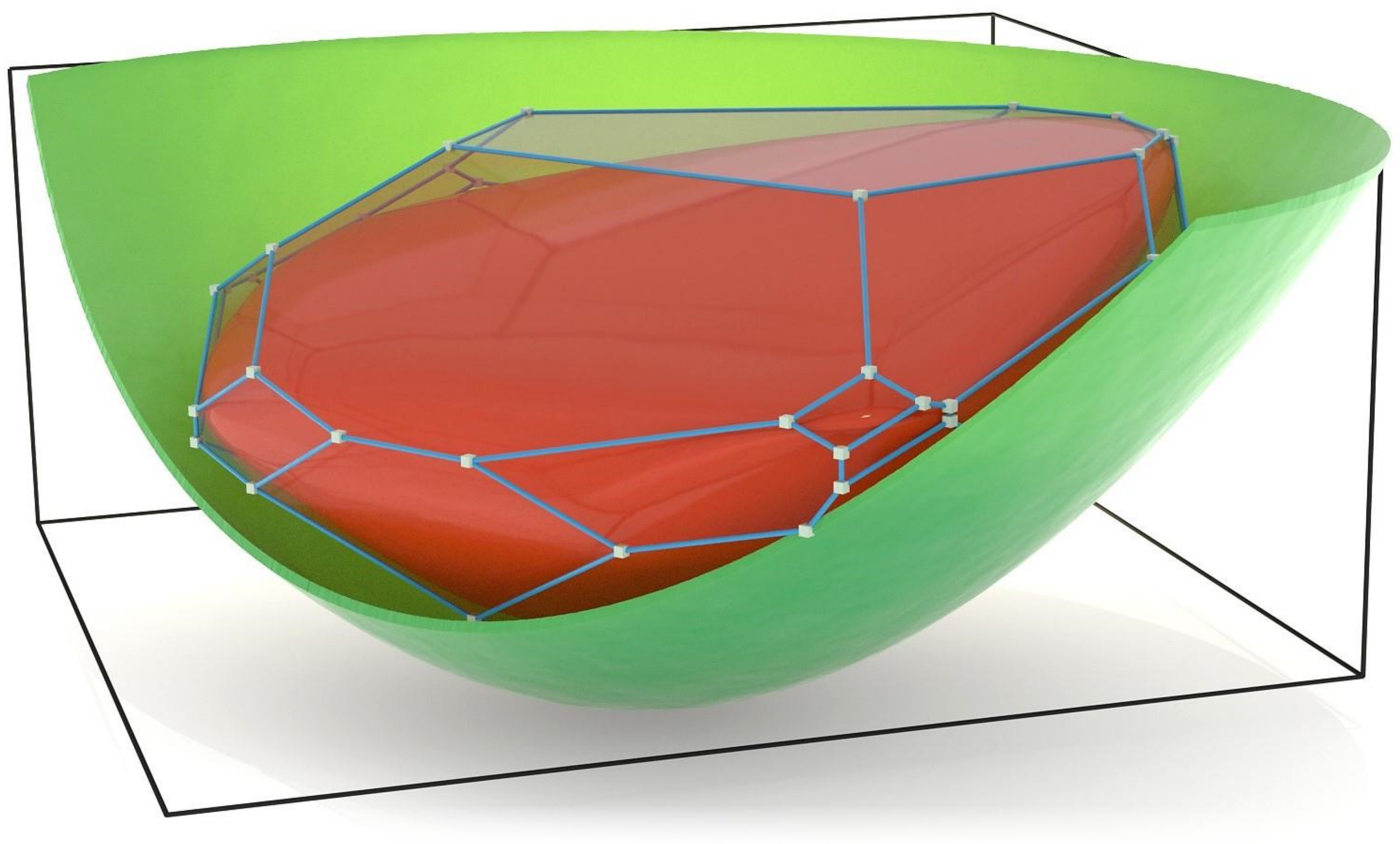}}%
    \put(0.57475612,0.03239445){\color[rgb]{0,0,0}\makebox(0,0)[lb]{
    
    \begin{rotate}{4}
    \smash{$\langle A\rangle$}
    \end{rotate}
    }}%
    \put(-0.010868775,0.08060685){\color[rgb]{0,0,0}\makebox(0,0)[lb]{
      \begin{rotate}{1}
    \smash{$\langle B \rangle$}
    \end{rotate} 
    }}%
    
    \put(-0.08,0.571){\color[rgb]{0,0,0}\makebox(0,0)[lb]{
        \begin{rotate}{1}
   \smash{$\langle A^2+B^2\rangle$}
    \end{rotate}
    }}%
  \end{picture}%
\endgroup%

	\caption{\label{fig:pott} Minimizing the sum of the variances of two observables $A$ and $B$ can be expressed entirely in terms of the set $\C$ of
     possible triples
    $(\langle A \rangle_\rho,\langle B \rangle_\rho,\langle A^2+B^2\rangle_\rho)$ (red solid convex body), namely as finding that vertical displacement
    of the surface $z=x^2+y^2$ (green paraboloid) which just touches $\C$ from below. We successively approximate $\C$ by polytopes (blue edges, boxed vertices) from the outside, and perform the minimization on this polytope. This gives a converging sequence of correct state-independent uncertainty relations.\vspace{-0.6cm}}
\end{figure}
The problem of getting optimal uncertainty bounds becomes more difficult as the dimension $d$ of the Hilbert space increases. Indeed, naively it would seem to be a search problem on the $2d-2$ dimensional manifold of pure states, which in bad cases might scale exponentially with $d$. However, we can do much better. We reformulate the problem as a geometric problem in three dimensions, namely of getting a sequence of outer polyhedral approximation of a certain convex set, see Fig.~\ref{fig:pott}. Any such approximation gives a valid uncertainty bound. In the iteration step, i.e., for computing a tighter approximation, one has to compute the lowest eigenvalue of a certain hermitian combination of the operators $A$ and $B$.
Those eigenvalue problems now determine the scaling of our method as a function of dimension, which will be a low order polynomial in $d$. Moreover, if additional information is available about $A$ and $B$, for example, if they are both sparse in the same basis, eigenvalue computations can be speeded up considerably, and our method will speed up by the same factor.

Tight uncertainty bounds have only been obtained for a few specific pairs of observables.  One example is angular momentum measurement, where bounds for two or three orthogonal spin components \cite{Hofmann,AMU,He} are known. In those cases symmetry crucially helps to reduce the problem. Other examples are qubits \cite{Abbott}, for which the low dimension allows an analytical solution.

There are also variants, in which the sharpness of a distribution is measured by other quantities than the usual variance \cite{BLW,phasespace,numberangle,Huang}, for instance entropies \cite{MU,ourEnt}, or where more than two observables are considered simultaneously \cite{Weigert,Weigert2}. Quite different methods \cite{MUR} are needed for optimal measurement uncertainty relations \cite{BLW}, or information-disturbance bounds \cite{BLW}, so we will not consider these aspects here.

\section*{Methods \label{methods}}
\subsection*{Linear state independent bounds}\label{linbound}
Since we are interested in state-independent bounds \cite{Deutsch} we have no use for the often-cited general relation by Robertson \cite{Robertson} (and its improvements \cite{Maccone}), which have a state-dependent expression like $\exprho{i[A,B]}^2$, or similar, on the right hand side. Indeed, any relation of product form $\var\rho A\,\var\rho B\geq c$ is useless for state-independent relations in finite dimension: $A$ and $B$ have discrete eigenvalues, so the trivial $c=0$ is the best possible bound. We therefore consider bounds of the form
\begin{align}\label{UCR}
   \var\rho A + \var\rho B \geq c .
\end{align}
Here, $c$ is the largest constant for which the above holds on \textit{any} quantum state $\rho$.
Since our method handles arbitrary $A$ and $B$ we can also admit factors here, i.e., inequalities of the form $\alpha\var\rho A + \beta\var\rho B \geq c(\alpha,\beta)$. Each of these constrains the set of uncertainty pairs $(\var\rho A , \var\rho B)$ to a half-plane, and together they outline the uncertainty set (or, more precisely its ``lower convex hull'', see Fig.~\ref{fig:appl} and \cite{AMU,Abbott,MUR}).

To see the connection to eigenvalue problems we write the optimal constant in \eqref{UCR} as
\begin{equation}\label{c-saw}
  c=\min_\rho\min_{a,b}\left\langle(A-a\idty)^2+(B-b\idty)^2\right\rangle_\rho .
\end{equation}
Here we just wrote the variance as the minimal quadratic deviation, using that the minimum with respect to $a$ is attained at the expectation $a=\langle A \rangle_\rho$. On the other hand, if we fix $a$ and $b$, the minimization with respect to $\rho$ is exactly the ground state problem for the operator in parentheses.
 This suggested our previous ansatz \cite{AMU}, which we call the {\it see-saw} algorithm: One alternatingly minimizes with respect to $\rho$ and $(a,b)$. In many practical cases this converges quickly, and with the safeguard of trying out several initial values it seems fairly reliable. However, in general the method of Alternating Minimization may easily fail to find the global minimum, and there is no proof of convergence. Intermediate results of the see-saw algorithm give an upper bound on $c$, but as an upper bound on a lower bound this is useless for applications. Moreover, there are indications that the see-saw algorithm actually may get trapped.
\begin{figure}[t]
	\def\svgwidth{\linewidth}
\begingroup%
  \makeatletter%
  \providecommand\color[2][]{%
    \errmessage{(Inkscape) Color is used for the text in Inkscape, but the package 'color.sty' is not loaded}%
    \renewcommand\color[2][]{}%
  }%
  \providecommand\transparent[1]{%
    \errmessage{(Inkscape) Transparency is used (non-zero) for the text in Inkscape, but the package 'transparent.sty' is not loaded}%
    \renewcommand\transparent[1]{}%
  }%
  \providecommand\rotatebox[2]{#2}%
  \ifx\svgwidth\undefined%
    \setlength{\unitlength}{197.8bp}%
    \ifx\svgscale\undefined%
      \relax%
    \else%
      \setlength{\unitlength}{\unitlength * \real{\svgscale}}%
    \fi%
  \else%
    \setlength{\unitlength}{\svgwidth}%
  \fi%
  \global\let\svgwidth\undefined%
  \global\let\svgscale\undefined%
  \makeatother%
  \begin{picture}(1,0.44514661)%
    \put(0,0){\includegraphics[width=\unitlength]{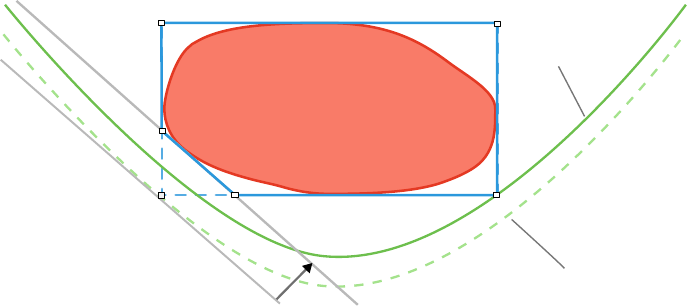}}%
    \put(0.34919608,0.28667639){\color[rgb]{0,0,0}\makebox(0,0)[lb]{\smash{\Large $\C$}}}%
    \put(0.4269864,0.01264312){\color[rgb]{0,0,0}\makebox(0,0)[lb]{\smash{$\r'$}}}%
    \put(0.76583789,0.02341306){\color[rgb]{0,0,0}\makebox(0,0)[lb]{\smash{
    \footnotesize $\mu(\x)=\mu(\v^*)$}}}%
    \put(0.14429084,0.12053265){\color[rgb]{0,0,0}\makebox(0,0)[lb]{\smash{$\v^*$}}}%
    \put(0.73463198,0.16635826){\color[rgb]{0,0,0}\makebox(0,0)[lb]{\smash{$\v^{**}$}}}%
    \put(0.74298256,0.35289528){\color[rgb]{0,0,0}\makebox(0,0)[lb]{\smash{\footnotesize$\mu(\x)=\mu(\v^{**})$}}}%
  \end{picture}%
\endgroup%
	\caption{\label{fig:2d}
    Two dimensional sketch of geometry and the basic algorithm: The set $\C$ (red) with its outer approximation $\P(\RR)$ (blue and blue dasehd) and the extremal points $\EE(\RR)$ (white squares). By adding the direction $\r'$, the polyhedral approximation is refined and the lower bound $c_-(\RR)$ is improved from   $\mu(\v^*)$ (dashed green parabola) to $\mu(\v^{**})$ (green parabola).}
\end{figure}
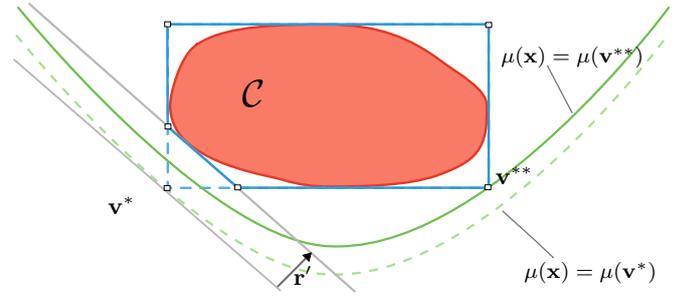

\subsection*{Geometry of outer approximations}
In contrast, the method described in this paper is an outer method, in which all intermediate steps give valid lower and upper bounds on $c$. Its geometric core is the joint numerical range
\begin{align}\label{numrange}
  \C=\left\{ \left( \exprho{A} ,\exprho{B},\exprho{A^2+B^2} \right) \Bigl\vert \;\rho \in \SS(\HH) \right\},
\end{align}
where $\SS(\HH)$ denotes the state space, i.e., the set of density operators.
Notice first that this set contains all the information necessary to compute $c$ from \eqref{c-saw}. With the quadratic functional $\mu(\x):=z-x^2-y^2$ of $\x=(x,y,z)\in\Rl^3$ we find
\begin{align}\label{deltaopt}
  c&=\min\limits_{\rho \in \SS(\HH)} \var\rho A + \var\rho B
  =\min\limits_{\x \in \C} \mu(\x).
\end{align}
Now the set $\C$ is clearly convex and compact, because the state space $\SS(\HH)$ has these properties, and they are preserved by the map taking $\rho$ to the tuple of expectations. The set $\C$ is therefore completely described by the linear inequalities it satisfies. To get such inequalities, let $\r=(r_1,r_2,r_3)$ be a real vector, and consider $H(\r)=r_1A+r_2B+r_3(A^2+B^2)$. Let $h(\r)$ denote the smallest eigenvalue of this operator. Then, for any state $\rho$, and hence the corresponding tuple $\x\in\C$ of expectations:
\begin{align}\label{lineq}
  \r\cdot\x=\exprho{H(\r)} \geq h(\r).
\end{align}

Now let $\RR\subset\Rl^3$ be any finite set of vectors, and consider the polytope $\P(\RR)$ of those points $\x$, which just satisfy the inequalities \eqref{lineq} with $\r\in\RR$. Since these vectors satisfy fewer constraints than $\C$, we have $\C\subset\P(\RR)$, i.e., this is an outer approximation of $\C$. Denote by $\EE(\RR)$ the set of extreme points of $\P(\RR)$, which is also finite. Then
\begin{equation}\label{clower}
  c\geq \min_{\x\in\P(\RR)} \mu(\x)=\min_{\x\in\EE(\RR)}\mu(\x)=:c_-(\RR).
\end{equation}
Here we have used, firstly, that the minimum over a larger set is smaller, and, secondly, that the functional $\mu$ is concave, so that the minimum over a compact convex set is attained at an extreme point.
Hence for every finite set $\RR$ of directions, we get a lower bound on $c$, which is computed as a finite minimum over $\EE(\RR)$. On the other hand, for each $r\in\RR$ we get a point $\x^*(r)$, with equality in Eq.~\eqref{lineq}. Then
\begin{equation}\label{cupper}
  c\leq \min_{\r\in\RR} \mu(\x^*(\r))=:c_+(\RR).
\end{equation}
So for every set $\RR$, this procedure estimates the optimal constant $c$ up to a precision $\varepsilon=c_+(\RR)-c_-(\RR)$.

\subsection{Basic algorithm}\label{algo}
\begin{figure}
\begin{minipage}[b]{\linewidth}
\begin{minipage}[l]{0.65\linewidth}
\includegraphics[width=.95\linewidth]{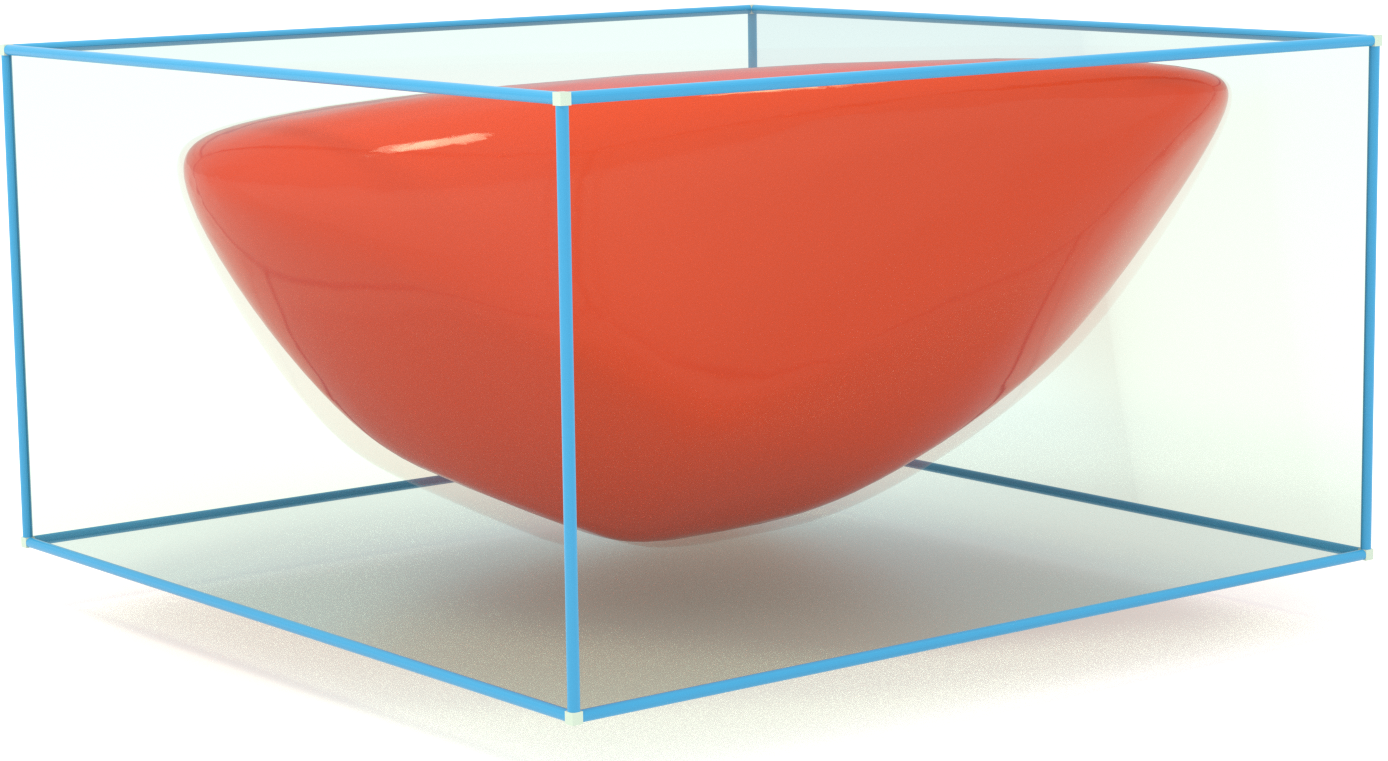}
\end{minipage}
\begin{minipage}[c]{0.2\linewidth}
\begin{align}
&{\rm steps}:&0      \nonumber\\
&{\rm vertices}:&8   \nonumber\\
&c_-(\RR):&-198.724  \nonumber\\
&\epsilon:&205.504   \nonumber
\end{align}
\end{minipage}

\end{minipage}

\begin{minipage}[b]{\linewidth}
\begin{minipage}[l]{0.65\linewidth}
\includegraphics[width=.95\linewidth]{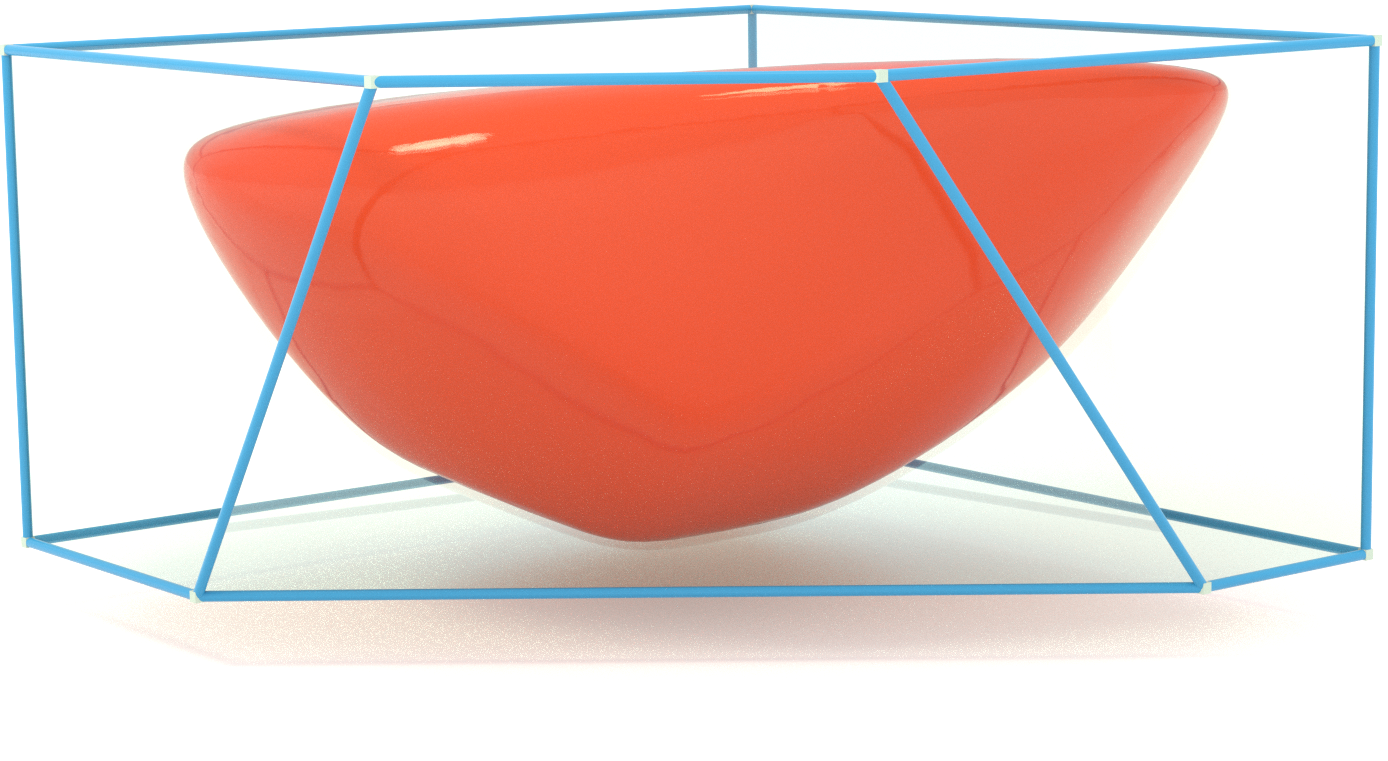}
\end{minipage}
\begin{minipage}[c]{0.2\linewidth}
\begin{align}
&{\rm steps}:&1      \nonumber\\
&{\rm vertices}:&10  \nonumber\\
&c_-(\RR):&-196.176  \nonumber\\
&\epsilon:&202.956   \nonumber
\end{align}
\end{minipage}

\end{minipage}

\begin{minipage}[b]{\linewidth}
\begin{minipage}[l]{0.65\linewidth}
\includegraphics[width=.95\linewidth]{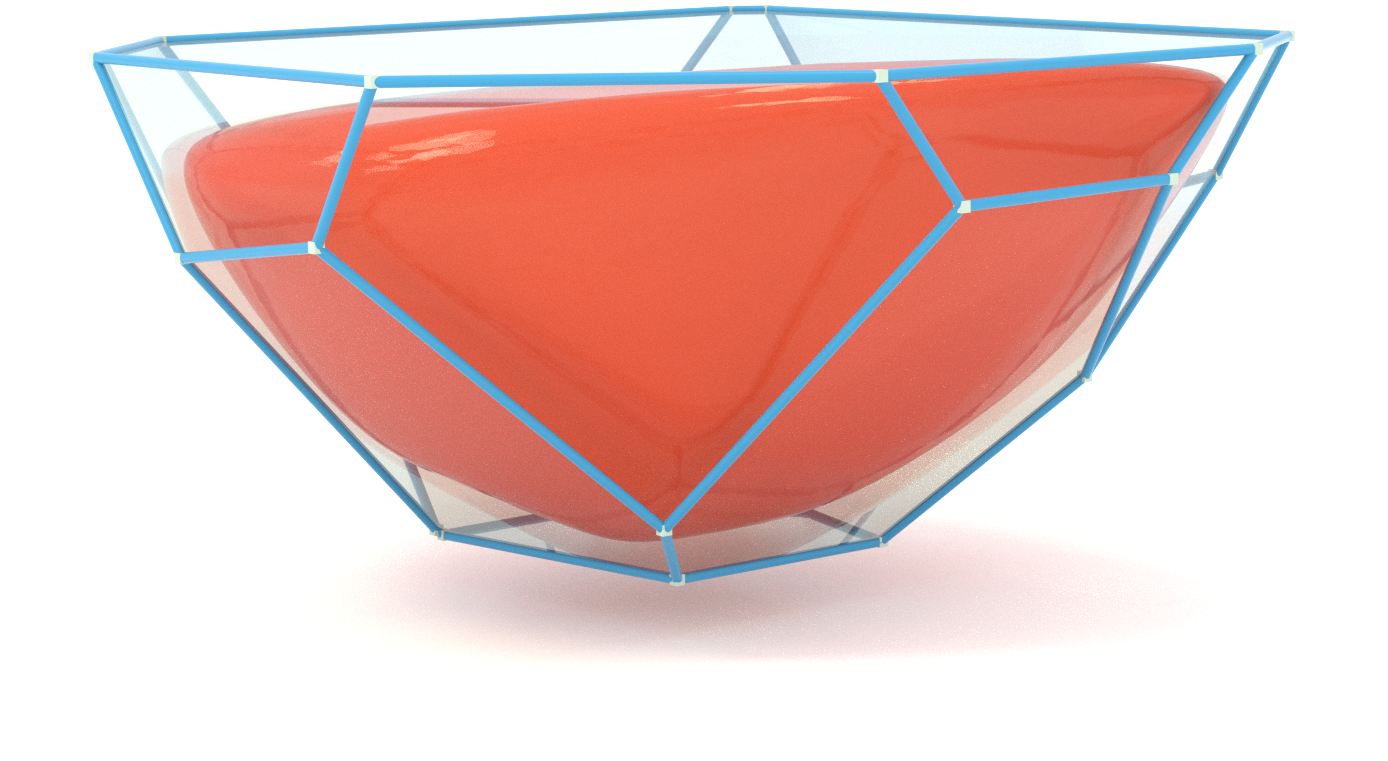}
\end{minipage}
\begin{minipage}[c]{0.2\linewidth}
\begin{align}
&{\rm steps}:&10     \nonumber\\
&{\rm vertices}:&28   \nonumber\\
&c_-(\RR):&-15.948\nonumber\\
&\epsilon:&22.724 \nonumber
\end{align}
\end{minipage}

\end{minipage}

\begin{minipage}[b]{\linewidth}
\begin{minipage}[l]{0.65\linewidth}
\includegraphics[width=.95\linewidth]{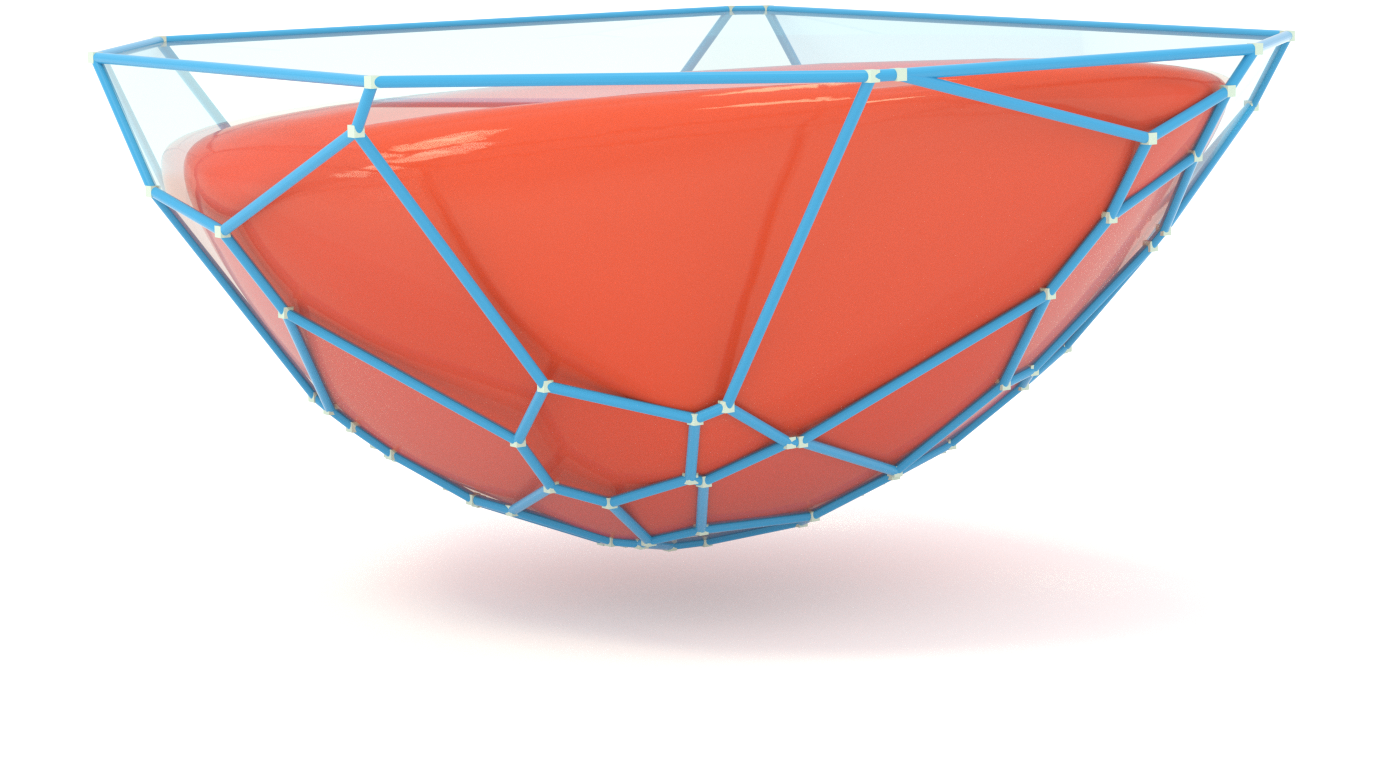}
\end{minipage}
\begin{minipage}[c]{0.2\linewidth}
\begin{align}
&{\rm steps}:&63      \nonumber\\
&{\rm vertices}:&132  \nonumber\\
&c_-(\RR):&\quad 6.629      \nonumber\\
&\epsilon:&0.007       \nonumber
\end{align}
\end{minipage}
\end{minipage}
\caption{\label{fig:algo} Improving the outer approximation of $\C$ (red convex body) by adding more directions to the set $\RR$. Every direction $\r\in\RR$ gives a face of $\P(\RR)$ (blue polytope). New directions are chosen such that the vertex with the lowest value of $\mu$ will be cut off. Example generated from randomly chosen $A,B\in\mathbb{R}^{10\times 10}$.}
\end{figure}

The idea of the algorithm is now to let the set $\RR$ grow step by step, which shrinks $\P(\RR)$, so $c_-(\RR)$ increases and $c_+(\RR)$ decreases (see Fig.~\ref{fig:2d} and Fig.~\ref{fig:algo}). The algorithm terminates when $\varepsilon$ is below the target accuracy.

Apart from the set $\RR$ it is useful to keep track of the polytope $\P(\RR)$ in the form of a list of vertices $\EE(\RR)$ and edges. To arrive at the next approximation $\RR'=\RR\cup\{\r'\}$:
\begin{enumerate}
  \item Determine a vertex $\v^*\in\EE(\RR)$ at which $\mu$ becomes minimal, and set
    \begin{equation}\label{newdir}
       \r'=\nabla\mu|_{\v^*}.
    \end{equation}
  \item Solve the minimum-eigenvalue problem for $H(\r')$. This provides the bound $h(\r')$ for the new inequality \eqref{lineq},
    and an expectation tuple $\x^*$ corresponding to the ground state.
  \item Compute $\mu(\x^*)$ and update $c_+(\RR')$, if this is smaller than the current value.
  \item Take the new inequality \eqref{lineq}, and compute the intersections with all current edges of $\P(\RR)$. This will give some new extreme points for $\EE(\RR')$, and corresponding edges.
  \item Evaluate $\mu$ on the new extreme points in $\EE(\RR')$ and update $c_-(\RR')$.
  Terminate if $c_+(\RR')-c_-(\RR')$ is as small as desired. Otherwise go to step 1.
\end{enumerate}

All these steps except the choice in step 1 are dictated by the geometry of outer approximation. The rationale of the choice \eqref{newdir} (apart from its flavour of gradient search) is that, whenever possible, it will eliminate the vertex $\v^*$ from $\P(\RR')$, and thus strictly increase $c_-(\RR)$, unless there are other vertices with the same value of $\mu$, which have first to be eliminated in a similar manner. A proof of this statement is provided in the appendix. As an application of our method, we derived the uncertainty relations for two non-orthogonal spin components, see the appendix.

\subsection{Generalization to POVMs}
Our method can be applied with minimal modifications to generalized measurements, i.e. observables given by positive operator valued measures (POVMs).
In general, a POVM measurement $\mathcal{A}$ is described by its outcomes $\{a_i\}$ and corresponding effects $\{E_i\}$  \cite{Ludwig,Heinosaari}, where the probability of obtaining the outcome $a_i\in\Rl$ is given by $\tr(\rho E_i)$. The moments of an outcome distribution are then given by the expectations of the moment operators $A^{(n)}=\sum_i (a_i)^n E_i$. The only difference from the ``standard'' projection valued case is that the identity $A^{(n)}=\bigl(A^{(1)}\bigr)^n$ no longer holds. But this is not required for our method.

We therefore only need to express variances as $\var{\rho}{A}=\langle A^{(2)}\rangle_\rho-\langle A^{(1)}\rangle_\rho^2$, and replace in \eqref{numrange} and the definition of $H(\r)$: $A^2$ by $A^{(2)}$, $A$ by $A^{(1)}$, and analogously for $B$.

\section*{Application to entanglement detection\label{appl}}
In \cite{Hofmann,Guehne}, it was shown that every state-independent uncertainty relation like \eqref{deltaopt} yields a non-linear entanglement witness, when applied to local measurements in a bipartition.
Here the following scenario is considered: Two parties, Alice and Bob, can  perform local measurements $A_1,A_2$ such as $B_1,B_2$, on an unknown quantum state $\rho$. Their goal is to decide if $\rho$ is entangled or not.
For this, they measure the 'sum observables' $M_1,M_2$, given by
\begin{align}
M_i=A_i \otimes \idty + \idty \otimes B_i.
\end{align}
In the POVM case this is generalized to measuring $A_i$ on Alice's side, $B_i$ on Bob's, and adding the outcomes, which results in
\begin{eqnarray}\label{povmSum}
  M_i^{(1)} &=& A_i^{(1)} \otimes \idty + \idty \otimes B_i^{(1)} \\
  M_i^{(2)} &=& A_i^{(2)} \otimes \idty + 2 A_i^{(1)} \otimes B_i^{(1)} +\idty \otimes B_i^{(2)}.
\end{eqnarray}
Now if $\rho=\rho_A\otimes\rho_B$ is uncorrelated, variances just add up, so
\begin{equation}\label{sepIneq}
   \var{\rho}{M_1}+\var{\rho}{M_2}\geq c_A+c_B,
\end{equation}
where $c_A$ and $c_B$ are the optimal uncertainty constants for the observable pairs $(A_1,A_2)$ and $(B_1,B_2)$, respectively.
Since the variance is concave, this inequality holds also for all convex combinations of uncorrelated states, i.e., for all separable states \cite{Hofmann}.

Hence if \eqref{sepIneq} is violated, $\rho$ must be entangled. Of course, there is also an uncertainty bound $c_M$ for the observable pair $(M_1,M_2)$. So the interesting range allowing the conclusion ``$\rho$ is entangled'' is marked by
\begin{equation}\label{entDetect}
  c_A+c_B>\var{\rho}{M_1}+\var{\rho}{M_2}\geq c_M.
\end{equation}

For angular momentum measurements, \eqref{sepIneq} can be seen \cite{ssreview} as a spin-squeezing criterion. As such, it requires the same experimental data as other spin squeezing criteria, see \cite{Duan-Zoller, Soerensen-Moelmer}, namely only a measurement of first and second moments of the total angular momentum. In contrast to entanglement criteria based on single outcomes, this requirement is very advantageous in typical experimental implementation, especially including many particle systems, see \cite{Klempt}.

We further sharpen this criterion by applying it to the observable pairs $(\mu A_1,\lambda A_2)$ and $(\mu B_1,\lambda B_2)$. In this way we get two convex regions of pairs $(\var{\rho}{M_1},\var{\rho}{M_2})$: A larger one containing the pairs achievable with arbitrary states, given by the bounds of the type $c_M$, and a smaller one attainable by separable states, given by the bounds of the type $c_A+c_B$. As Fig.~\ref{fig:appl} shows, this increases the parameter range for which entanglement can be certified. The linear uncertainty bound with equal weights as a function of the local noise, evaluated for measurements $M_1$ and $M_2$ on separable and entangled states is shown by Fig.~\ref{fig:lubed}.

\subsection*{Entanglement detection with noisy detectors}

\begin{figure}[b]
\def\svgwidth{0.9\linewidth}
\begingroup%
  \makeatletter%
  \providecommand\color[2][]{%
    \errmessage{(Inkscape) Color is used for the text in Inkscape, but the package 'color.sty' is not loaded}%
    \renewcommand\color[2][]{}%
  }%
  \providecommand\transparent[1]{%
    \errmessage{(Inkscape) Transparency is used (non-zero) for the text in Inkscape, but the package 'transparent.sty' is not loaded}%
    \renewcommand\transparent[1]{}%
  }%
  \providecommand\rotatebox[2]{#2}%
  \ifx\svgwidth\undefined%
    \setlength{\unitlength}{539bp}%
    \ifx\svgscale\undefined%
      \relax%
    \else%
      \setlength{\unitlength}{\unitlength * \real{\svgscale}}%
    \fi%
  \else%
    \setlength{\unitlength}{\svgwidth}%
  \fi%
  \global\let\svgwidth\undefined%
  \global\let\svgscale\undefined%
  \makeatother%
  \begin{picture}(1,0.98886827)%
    \put(0,0){\includegraphics[width=\unitlength]{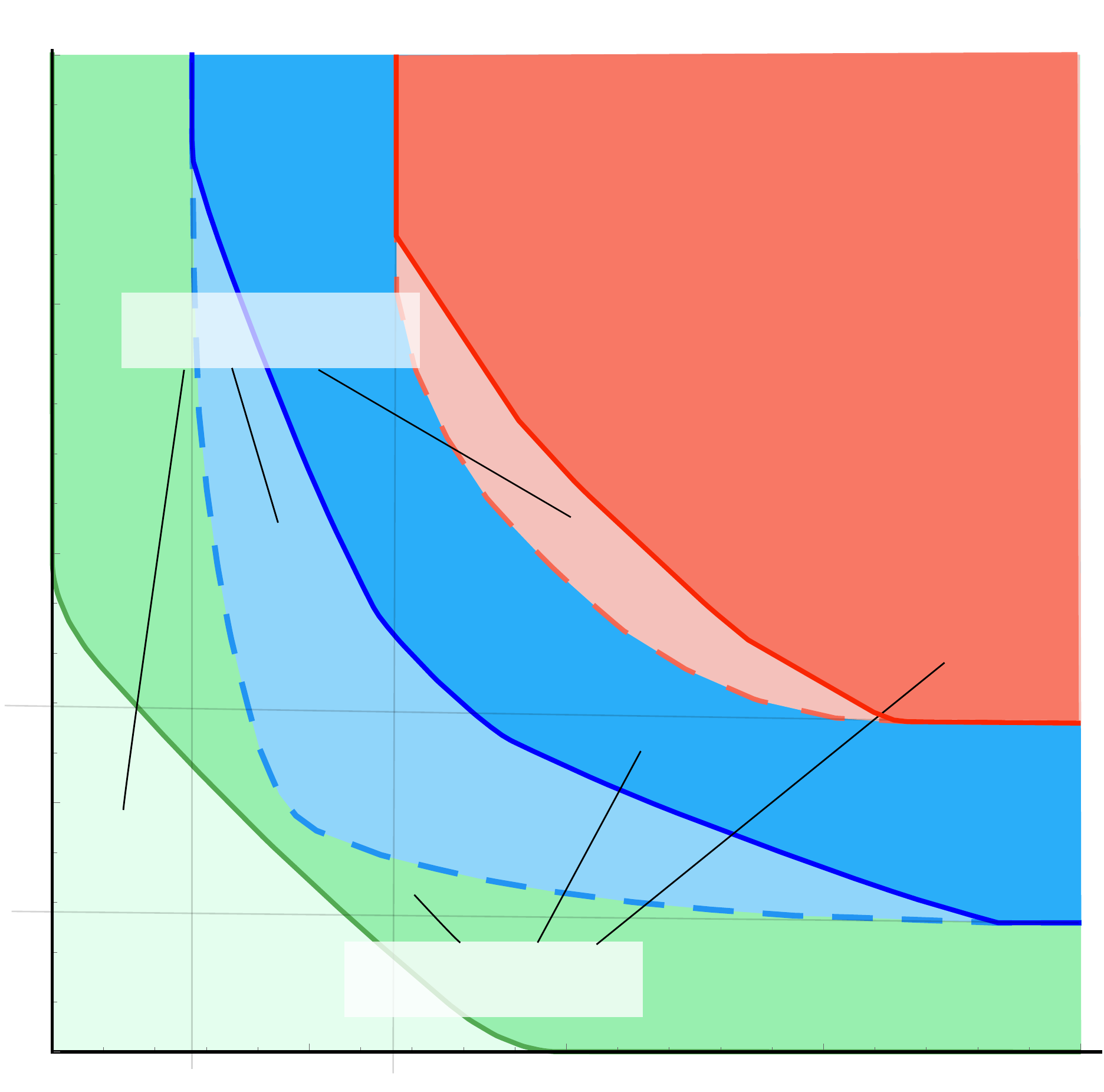}}%
    \put(-0.01887311,0.90537198){\color[rgb]{0,0,0}\makebox(0,0)[lb]{\smash{$2$}}}%
    \put(-0.05308063,0.69712535){\color[rgb]{0,0,0}\makebox(0,0)[lb]{\smash{$1.5$}}}%
    \put(-0.01533441,0.47359703){\color[rgb]{0,0,0}\makebox(0,0)[lb]{\smash{$1$}}}%
    \put(-0.05367038,0.24829928){\color[rgb]{0,0,0}\makebox(0,0)[lb]{\smash{$0.5$}}}%
    \put(0.79404706,0.07407154){\color[rgb]{0,0,0}\makebox(0,0)[lb]{\smash{$\alpha=0$}}}%
    \put(0.7814549,0.25110254){\color[rgb]{0,0,0}\makebox(0,0)[lb]{\smash{$\alpha=0.2$}}}%
    \put(0.75997415,0.65627398){\color[rgb]{0,0,0}\makebox(0,0)[lb]{\smash{$\alpha=0.5$}}}%
    \put(0.13799229,0.67476701){\color[rgb]{0,0,0}\makebox(0,0)[lb]{\smash{ entangled}}}%
    \put(0.33904936,0.08769382){\color[rgb]{0,0,0}\makebox(0,0)[lb]{\smash{ separable}}}%
    \put(0.94705229,-0.01){\color[rgb]{0,0,0}\makebox(0,0)[lb]{\smash{$2$}}}%
    \put(0.70932896,-0.01){\color[rgb]{0,0,0}\makebox(0,0)[lb]{\smash{$1.5$}}}%
    \put(0.48767588,-0.01){\color[rgb]{0,0,0}\makebox(0,0)[lb]{\smash{$1$}}}%
    \put(0.23843387,-0.01){\color[rgb]{0,0,0}\makebox(0,0)[lb]{\smash{$0.5$}}}%
    \put(-0.01668162,0.96002761){\color[rgb]{0,0,0}\makebox(0,0)[lb]{\smash{$\var{\rho}{M_1^\alpha}$}}}%
    \put(1.01,0.03572051){\color[rgb]{0,0,0}\makebox(0,0)[lb]{\smash{$\var{\rho}{M_2^\alpha}$}}}%
  \end{picture}%
\endgroup%

\caption{\label{fig:appl}
Uncertainty regions for entangled and separable states. Superposition of the graphs for different noise levels $\alpha$: green$=0$, blue$=0.2$, red$=0.5$. In this example we consider local measurements of orthogonal spin-1 components, i.e. $M_i=L_i^A+L_i^B$.\\
}
\end{figure}
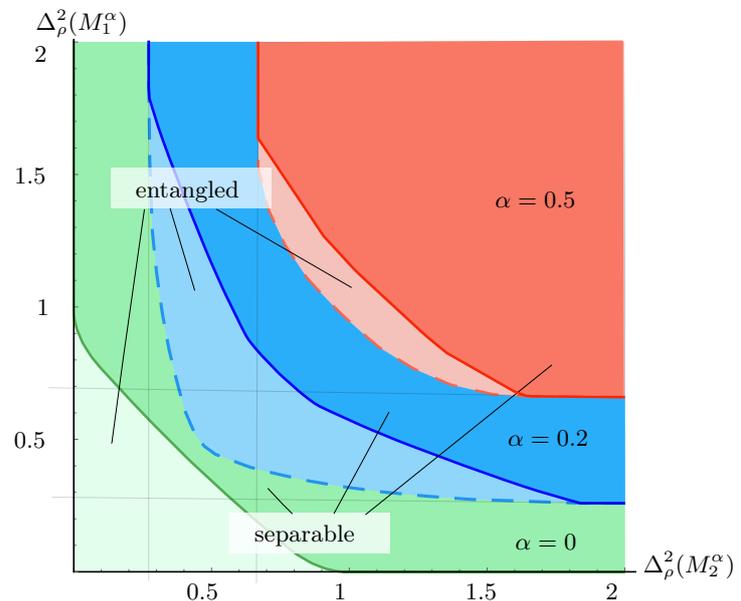
The generalization to POVMs increases the possibilities for entanglement detection. Suppose for the sake of discussion that before hitting the detector each subsystem goes through a known noisy channel. This typically increases variance \cite{Lucke}, so traditional spin squeezing inequalities would often fail to detect entanglement. Indeed the state after the action of the noisy channels may well fail to be entangled. On the other hand, we might be interested in the presence of entanglement {\it before} the action of the noise. This is the appropriate view when the noise is inherent in the detection process. The noise is thus applied in the Heisenberg picture, turning even a standard projection valued measurement into a proper POVM. This might easily find entanglement, which would go undetected by a direct application of the spin squeezing criterion.

These possibilities are shown in Fig.~\ref{fig:appl} by superimposing the entanglement detection regions for three different noise levels of a partially depolarizing channel $\rho\mapsto (1-\alpha)\rho+\alpha \rho_0$, where $\rho_0\propto\idty$ is the maximally mixed state, and $\alpha$ is a noise parameter. Increasing $\alpha$ shifts the diagram towards larger variances, but even for a modest noise level of $\alpha=0.2$ the entanglement detection region lies entirely in the region where traditional spin squeezing (corresponding to $\alpha=0$) would never find any entanglement.

\section*{Conclusions and Outlook}
We provided an algorithm for determining the optimal uncertainty bounds for two arbitrary observables. The precision of the bound is controlled as a duality gap, so terminating the iteration at any step gives a certified lower uncertainty bound together with an error estimate.

The method can, in principle, be extended to more observables, or to variances based not on quadratic but higher order deviations. However, this would increase the dimension of the geometric problem. Thus at every new approximation step one has to determine the intersection of the polytope with the new supporting hyperplane. This requires a better book-keeping of the topological structure of the polytopes, and a local version of the vertex enumeration problem \cite{Avis}.

The inequalities derived here have an immediate application to entanglement detection by generalized spin squeezing criteria. The possibility to use arbitrary observables (rather than orthogonal angular momentum components) greatly increases the versatility of this method.

It is an apparently open problem how strong the method becomes with arbitrary $A_i$, $B_j$, i.e. is every entangled state violating a local uncertainty relation. The problem has been studied carefully for orthogonal spin components \cite{Guehne,Guehne2}, but we do not know of a characterization of the (un-)detectable, possibly entangled states.

\begin{acknowledgments}
We gratefully acknowledge inspiring conversations and email exchange with Marcus Cramer, Otfried G\"uhne, G{\'e}za T{\'o}th, Kais Abdelkhalek, David Reeb and Terry Farrelly. \\
We also acknowledge the financial support from the RTG 1991 and CRC 1227 DQ-mat funded by the DFG and the
collaborative research project Q.com-Q funded by the BMBF.
\end{acknowledgments}

\bibliographystyle{unsrt}
\bibliography{library}

\newpage\quad\newpage
\appendix
\section{Appendix}
\subsection{Strict monotonicity of the gap}
\label{convergence}
We will consider measurements $A$ and $B$ which could also be represented by two general POVMs. Furthermore, we will assume that we already have an initial outer approximation of the corresponding set $\C$ by a polyhedron $\P(\RR_0)$, constructed from initial directions $\RR_0$.
Such a set of directions can be constructed by taking the face normals of a cube, as in Fig.~\ref{fig:algo}.

Let $\v^*$ be a vertex of $\EE(\RR)$ on which the minimum of $\mu$ is attained, i.e. $c_-(\RR)=\mu(v^*)$ and take $\r'=\nabla\mu|_{\v^*}$ as new direction such as $\RR':=\RR\cup\r'$ as new set of directions, in every step. Then the bound $c_-(\RR)$ will either increase after a finite round of such steps or attain a global minimum on $\C$.

\begin{proof}
We will show that, by taking $\r'$ as above, the point $\v^*$ will be removed from the resulting polyhedron $\PP(\RR')$ whenever $\v^*$ is not in $\C$.
From this statement we can conclude that: If $\v^*$ is removed, and there is no point in $\C$ which attains the value $c_-(\RR')$, the new bound $c_-(\RR')$ will fail to increase if and only if there is another extremal point in $\EE(\RR')$ that also attains the same minimal value of $\mu$. However, because $\EE(\RR')$ is a finite set, all those points will be removed from it after a finite round of steps. Hence, the bound $c_- $ will increase after a finite round of steps.

In the alternative case,  when $\v^*$ is in $\C$, the upper bound $c_+$ will also be attained on $\v^*$, such that $c_+=c_-=c$. Therefore, we already would have found the optimal bound.

Consider a fixed $\v^*$ and the level-set $M_{\v*}=\{\x\in\mathbb{R}^3|\mu(\x)\geq\mu(\v^*)\}$. The set $M_{\v*}$ is convex and obviously contains $\C$ and $\P(\RR)$ as subsets. $\mu$ is a quadratic functional, so its gradient is well defined everywhere. Moreover, the direction $\r'=\nabla\mu|_{\v^*}$  is the normal direction of the tangent space of $M_{\v^*}$ at the point $\v^*$. Due to convexity, $M_{\v^*}$ is described by linear inequalities corresponding to its tangent spaces, which implies that
\begin{align}
\r'.\x\geq\r'.\v^*
\end{align}
for all points $\x$ from $M_{\v^*}$, and so, for all $\x$ from $\C$ and $\P(\RR)$, as well. More precisely, we have $\r'.\v^*=\min \{\r'.\x|\x \in M_{\v^*}\}$, with a minimum that is attained uniquely on $\v^*$, because $\mu$ is strictly concave. If we now consider the new set of direction $\RR'$ and its corresponding inequalities for constructing $\P(\RR')$, see \eqref{lineq}, we  have
\begin{align}
\r'.\x\geq h(\r')=\min \{\r'.\x|\x \in \C \} \geq \r'.\v^* \label{label1}
\end{align}
for all $\x\in\P(\RR')$. Hereby, equality in the last part of \eqref{label1}  holds if and only if $\v^*\in\C$. In all other cases the functional $\r'.\x$ separates the set $\P(\RR')$ from the point $\v^*$.
\end{proof}

\subsection{Precision per Step}
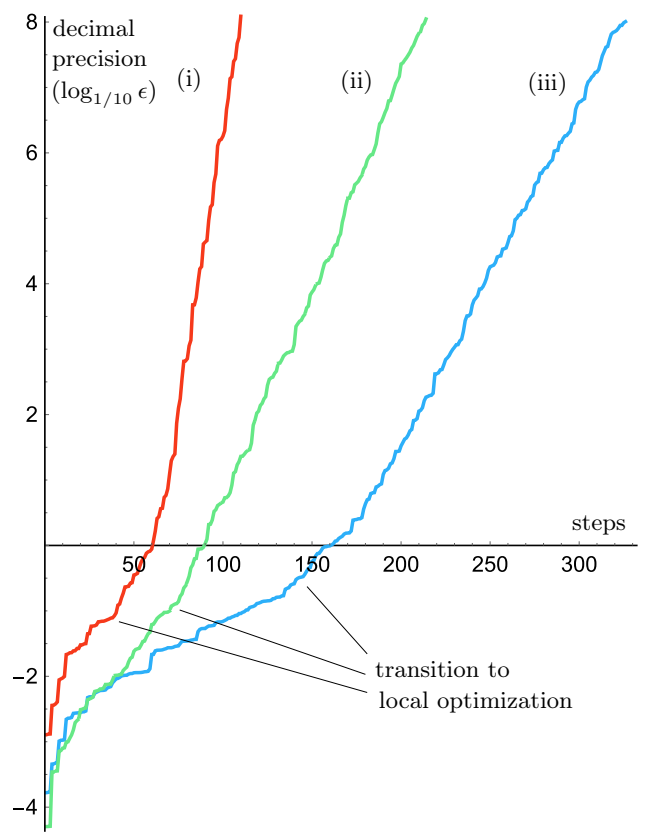
\begin{figure}[h!]
\def\svgwidth{0.95\linewidth}
\begingroup%
  \makeatletter%
  \providecommand\color[2][]{%
    \errmessage{(Inkscape) Color is used for the text in Inkscape, but the package 'color.sty' is not loaded}%
    \renewcommand\color[2][]{}%
  }%
  \providecommand\transparent[1]{%
    \errmessage{(Inkscape) Transparency is used (non-zero) for the text in Inkscape, but the package 'transparent.sty' is not loaded}%
    \renewcommand\transparent[1]{}%
  }%
  \providecommand\rotatebox[2]{#2}%
  \ifx\svgwidth\undefined%
    \setlength{\unitlength}{597.62109375bp}%
    \ifx\svgscale\undefined%
      \relax%
    \else%
      \setlength{\unitlength}{\unitlength * \real{\svgscale}}%
    \fi%
  \else%
    \setlength{\unitlength}{\svgwidth}%
  \fi%
  \global\let\svgwidth\undefined%
  \global\let\svgscale\undefined%
  \makeatother%
  \begin{picture}(1,1.3206852)%
    \put(0,0){\includegraphics[width=\unitlength]{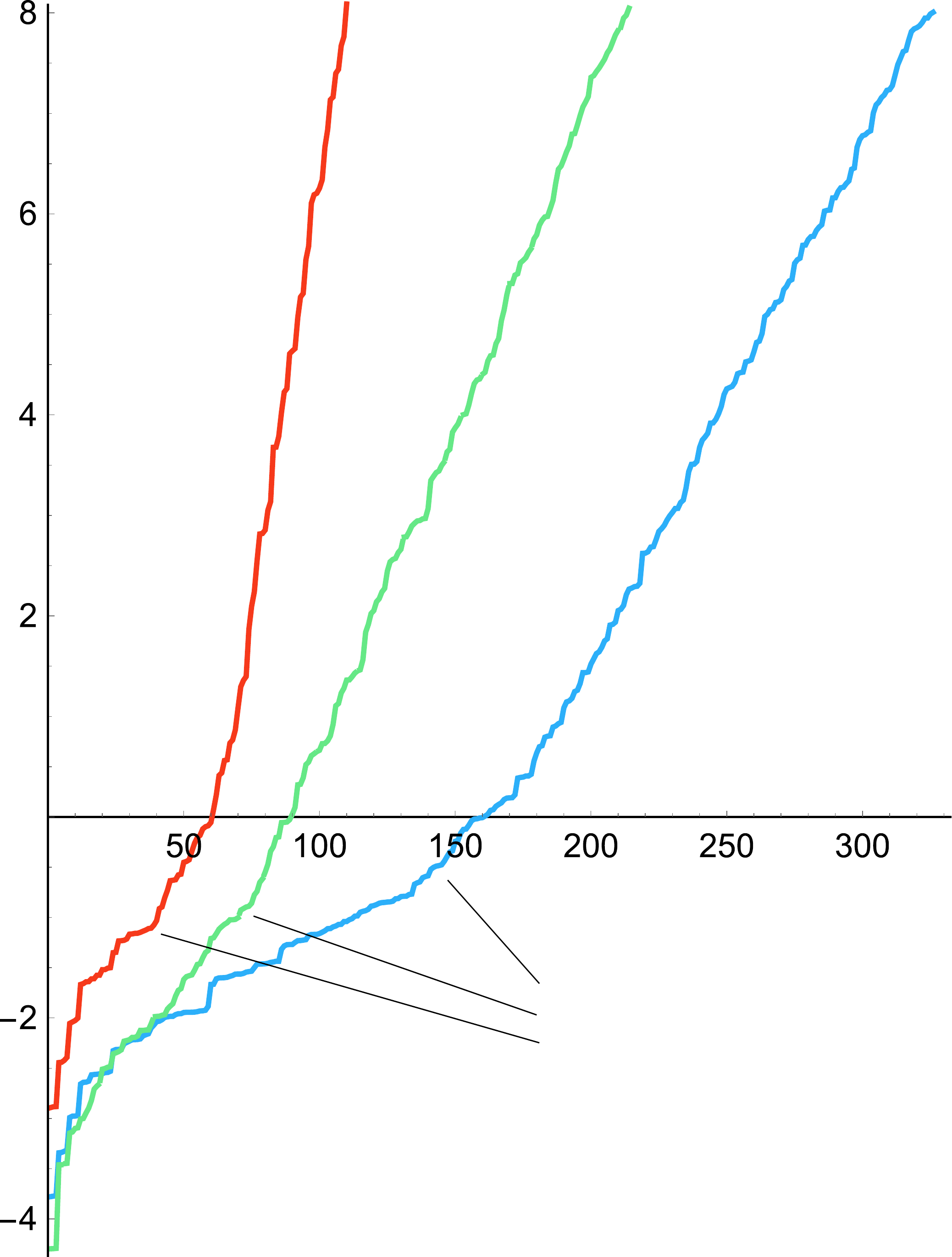}}%
    \put(0.89519505,0.48758269){\color[rgb]{0,0,0}\makebox(0,0)[lb]{\smash{steps}}}%
    \put(0.06308442,1.28478984){\color[rgb]{0,0,0}\makebox(0,0)[lb]{\smash{decimal 
}}}%
    \put(0.06234892,1.23564212){\color[rgb]{0,0,0}\makebox(0,0)[lb]{\smash{precision}}}%
    \put(0.26083745,1.20413756){\color[rgb]{0,0,0}\makebox(0,0)[lb]{\smash{(i)}}}%
    \put(0.52361613,1.20074689){\color[rgb]{0,0,0}\makebox(0,0)[lb]{\smash{(ii)}}}%
    \put(0.82369255,1.19735619){\color[rgb]{0,0,0}\makebox(0,0)[lb]{\smash{(iii)}}}%
    \put(0.06071127,1.18429081){\color[rgb]{0,0,0}\makebox(0,0)[lb]{\smash{($\log_{1/10}\epsilon$)}}}%
    \put(0.57965323,0.25054373){\color[rgb]{0,0,0}\makebox(0,0)[lb]{\smash{transition to 
}}}%
    \put(0.58570193,0.20027224){\color[rgb]{0,0,0}\makebox(0,0)[lb]{\smash{local optimization}}}%
  \end{picture}%
\endgroup%

\caption{Typical scaling behaviour of the gap $\epsilon$ in dependence of the steps made by the algorithm. Here depicted for three different randomly chosen pairs of operators, see Tab.~\ref{tab1}. In the above examples, the optimal uncertainty bound is attained at a unique point on $\C$, hence we observe a localization of our algorithm around this point. When this happens the scaling of the precision  $\epsilon$, measured in decimal places, becomes linear, i.e. $\epsilon\approx 10^{-\lambda \# \text{steps}}$ .
\label{num1}}
\end{figure}

A crucial property of any numerical method is its performance. For the method provided in this work we measure it by the numerical precision in comparison to the number of steps required. As a benchmarking we computed several random examples and illustrated three of them in  Fig.~\ref{num1}. We observe that in the typical case, there are two different kinds of scaling behaviour. During the first part of steps, the precision increases slower than in the second. In the regime of the first steps the algorithm improves the outer approximation at very different points. However, once the outer polyherdron is fine enough, the algorithm generates vertices close to the actual optimum. If this optimum is unique, the algorithm will go to a regime where all improvements are made locally.  In Fig.~\ref{num1} this transition from global to local optimization is marked. After this point the improvement of precision per step, measured in decimal places of the gap $\epsilon$, scales linear. 

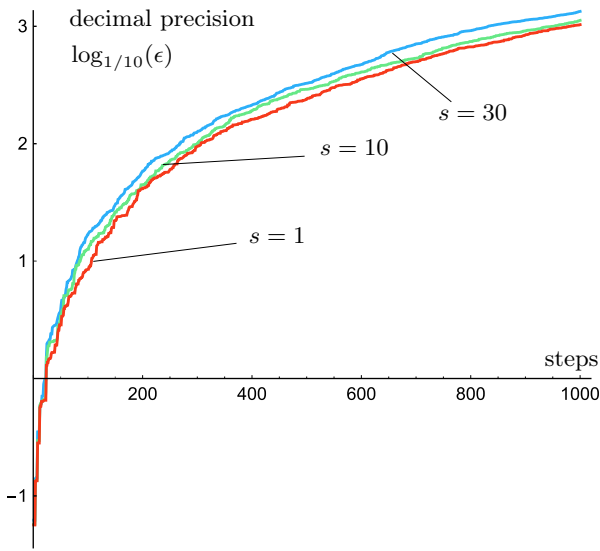
\begin{figure}[h]
\def\svgwidth{0.9\linewidth}
\begingroup%
  \makeatletter%
  \providecommand\color[2][]{%
    \errmessage{(Inkscape) Color is used for the text in Inkscape, but the package 'color.sty' is not loaded}%
    \renewcommand\color[2][]{}%
  }%
  \providecommand\transparent[1]{%
    \errmessage{(Inkscape) Transparency is used (non-zero) for the text in Inkscape, but the package 'transparent.sty' is not loaded}%
    \renewcommand\transparent[1]{}%
  }%
  \providecommand\rotatebox[2]{#2}%
  \ifx\svgwidth\undefined%
    \setlength{\unitlength}{695.84858398bp}%
    \ifx\svgscale\undefined%
      \relax%
    \else%
      \setlength{\unitlength}{\unitlength * \real{\svgscale}}%
    \fi%
  \else%
    \setlength{\unitlength}{\svgwidth}%
  \fi%
  \global\let\svgwidth\undefined%
  \global\let\svgscale\undefined%
  \makeatother%
  \begin{picture}(1,0.91803377)%
    \put(0,0){\includegraphics[width=\unitlength]{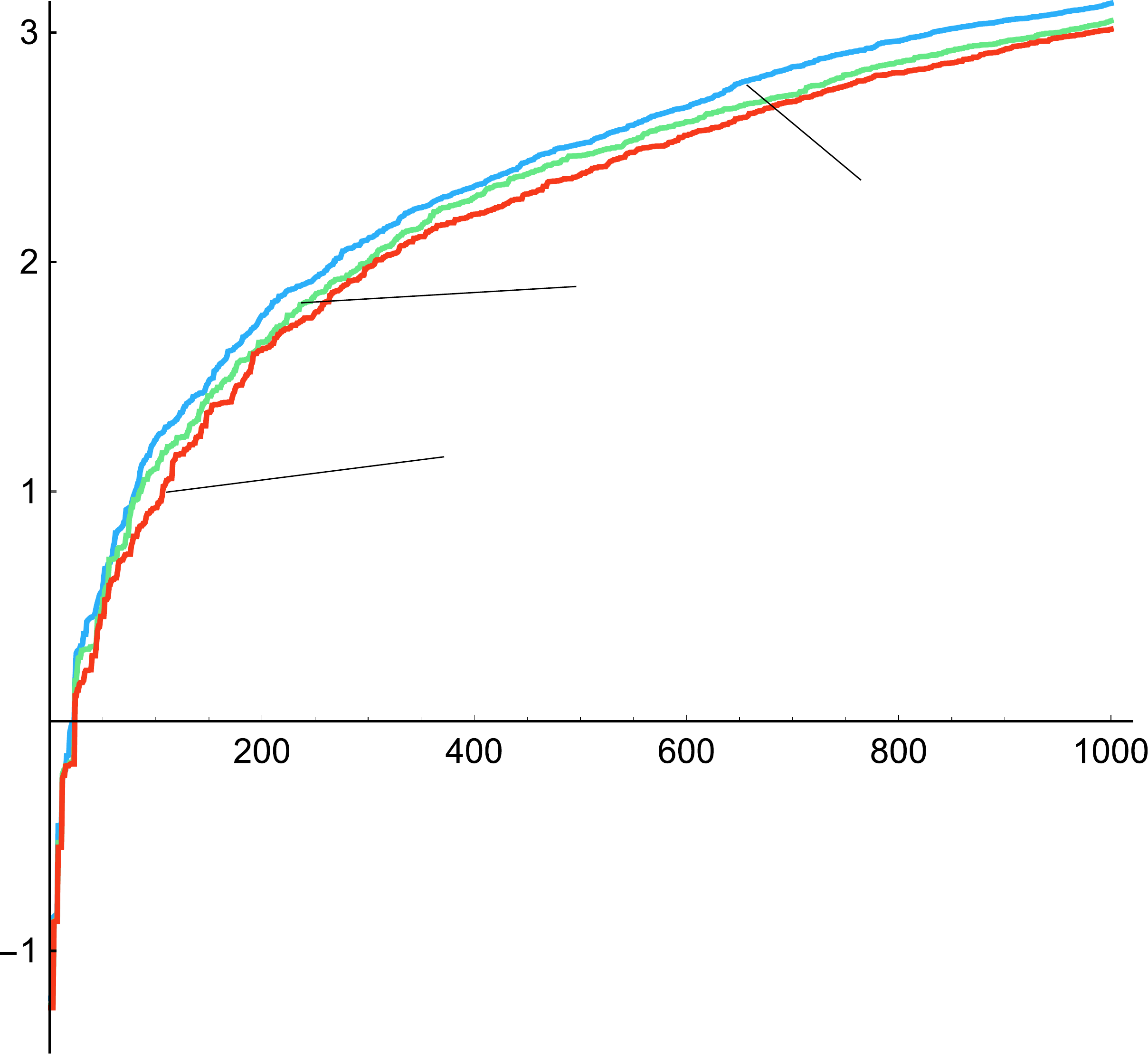}}%
    \put(0.07692857,0.88969483){\color[rgb]{0,0,0}\makebox(0,0)[lb]{
    \smash{
    decimal precision  
   }}}%
   
    \put(0.07992857,0.82969483){\color[rgb]{0,0,0}\makebox(0,0)[lb]{
    \smash{
   $\log_{1/10}(\epsilon)$ 
   }}}%
   
    \put(0.38,0.52){\color[rgb]{0,0,0}\makebox(0,0)[lb]{
    \smash{
   $s=1$ 
   }}}%
   \put(0.5,0.67){\color[rgb]{0,0,0}\makebox(0,0)[lb]{
    \smash{
   $s=10$ 
   }}}%
   \put(0.7,0.73){\color[rgb]{0,0,0}\makebox(0,0)[lb]{
    \smash{
   $s=30$ 
   }}}%
   
    \put(0.90825361,0.31050255){\color[rgb]{0,0,0}\makebox(0,0)[lb]{\smash{steps}}}%
  \end{picture}%
\endgroup%

	\caption{Scaling behaviour  of the precision $\epsilon$ for orthogonal angular momentum components. Depicted for different spins s: red=1, green =10, blue= 30. In these examples the optimal value of the uncertainty bound is not attained on a finite set of points, hence we have no localization of the algorithm. This benchmarks the worst case with respect to scaling. In the above example we rescaled the spectra of the operators to the unit interval. From the figure above can been seen that the algorithm shows the same scaling behaviour in all three cases. This illustrates that, the amount of steps the algorithm takes for reaching a certain target precision, is independent of the underlying Hilbert space dimension. 
 \label{num2}}
\end{figure}

\begin{table}[t]
\begin{tabular}{c|ccc}
sample&method&size&steps\\
\hline\\
(i)    &Haar random&$30\times30$&$110$\\
\\
(ii) &\makecell{\small Haar random eigenvectors\\ \small uniform dist. spectrum}&$200\times200$&$214$\\\\
(iii)   &Haar random&$200\times200$&$326$\\
\end{tabular}
\caption{Parameters of different random examples, see also Fig.-\ref{num1}, in order to benchmark the performance of the algorithm. 
\label{tab1}}
\end{table}
This behaviour agrees with the worst case example given in Fig.~\ref{num2}, where we considered two orthogonal angular momentum components, $L_z$ and $L_x$. Here rotations around the $y$-axis, in terms of spin components, impose a rotational degree of freedom on the set of all linear combinations of the operators $L_z$, $L_x$ and $L_x^2+L_z^2=s(s+1)-L_y^2$. This results in a region $\C$ that is rotational symmetric, as well. As the variance sum itself shares the same symmetry the optimum on $\C$ will be attained on a continuum of points. 

\enlargethispage{15\baselineskip}
This is  a \textit{worst case} scenario, because our method has to improve the outer approximation on a continuum of points. Hence, no localization of the algorithm can be expected, and no transition in a linear scaling regime happens, for this compare Fig.~\ref{num1} with Fig.~\ref{num2}. Note that, in this highly symetrical case there is no need to performe the algorithm on the whole set $\C$. If we take care of the ounderlying symmetry the problem reduces to a fast scaling problem on a two dimensioal subset of $\C$ again.

\subsection{Examples} \label{examples}
\begin{figure}[h!]
\def\svgwidth{0.9\linewidth}
\begingroup%
  \makeatletter%
  \providecommand\color[2][]{%
    \errmessage{(Inkscape) Color is used for the text in Inkscape, but the package 'color.sty' is not loaded}%
    \renewcommand\color[2][]{}%
  }%
  \providecommand\transparent[1]{%
    \errmessage{(Inkscape) Transparency is used (non-zero) for the text in Inkscape, but the package 'transparent.sty' is not loaded}%
    \renewcommand\transparent[1]{}%
  }%
  \providecommand\rotatebox[2]{#2}%
  \ifx\svgwidth\undefined%
    \setlength{\unitlength}{931bp}%
    \ifx\svgscale\undefined%
      \relax%
    \else%
      \setlength{\unitlength}{\unitlength * \real{\svgscale}}%
    \fi%
  \else%
    \setlength{\unitlength}{\svgwidth}%
  \fi%
  \global\let\svgwidth\undefined%
  \global\let\svgscale\undefined%
  \makeatother%
  \begin{picture}(1,0.68743287)%
    \put(0,0){\includegraphics[width=\unitlength]{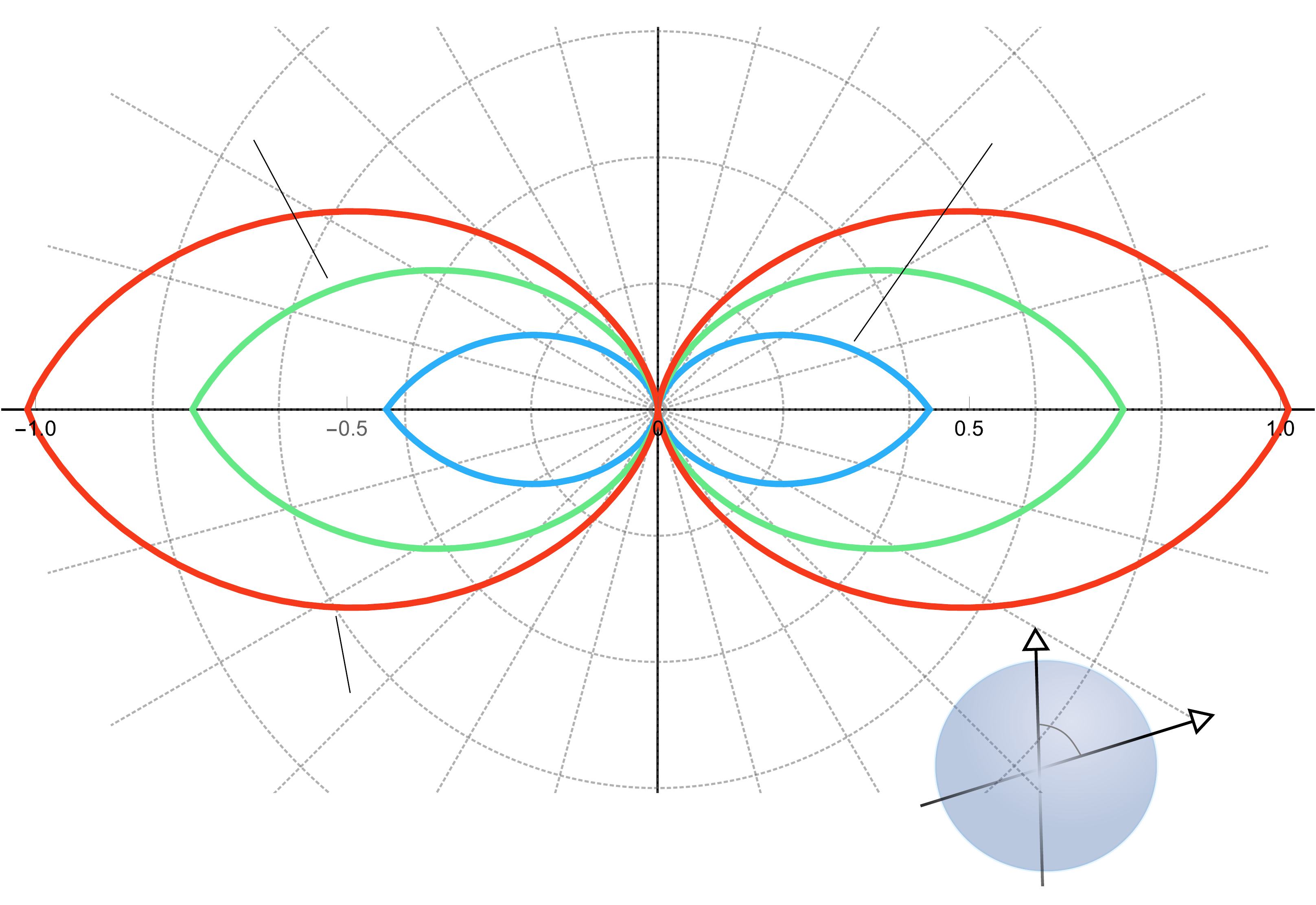}}%
    \put(0.5044376,0.57367066){\color[rgb]{0,0,0}\makebox(0,0)[lb]{\smash{0}}}%
    \put(0.49826637,0.18506194){\color[rgb]{0,0,0}\makebox(0,0)[lb]{\smash{$\pi$}}}%
    \put(0.24666016,0.14358235){\color[rgb]{0,0,0}\makebox(0,0)[lb]{\smash{s=3}}}%
    \put(0.15678466,0.5918865){\color[rgb]{0,0,0}\makebox(0,0)[lb]{\smash{s=2}}}%
    \put(0.7443078,0.58497385){\color[rgb]{0,0,0}\makebox(0,0)[lb]{\smash{s=1}}}%
    \put(0.31065167,0.64285139){\color[rgb]{0,0,0}\makebox(0,0)[lb]{\smash{$\phi$}}}%
    \put(0.7205073,0.38266958){\color[rgb]{0,0,0}\makebox(0,0)[lb]{\smash{$\Delta L_z+\Delta L_\phi$}}}%
    \put(0.79622288,0.18511951){\color[rgb]{0,0,0}\makebox(0,0)[lb]{\smash{$L_z$}}}%
    \put(0.87531688,0.09509481){\color[rgb]{0,0,0}\makebox(0,0)[lb]{\smash{$L_\phi$}}}%
    \put(0.79573155,0.12783106){\color[rgb]{0,0,0}\makebox(0,0)[lb]{\smash{$\phi$}}}%
  \end{picture}%
\endgroup%

\caption{Polar plot of the uncertainty bound between the non-orthogonal spin components $L_z$ and $L_\phi$ as a function of the angle $\phi$, depicted for spins $s=1,2,3$ \label{angularplot}}
\end{figure}
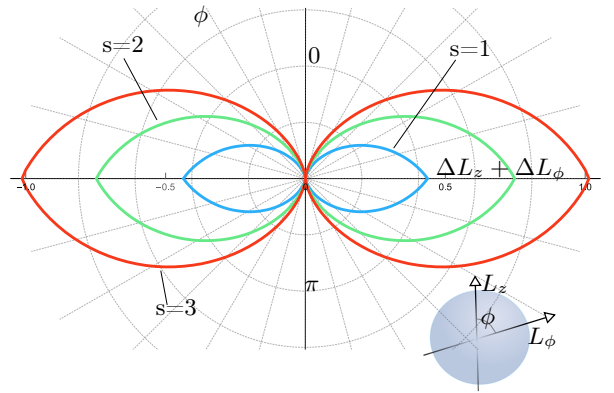

{\bf Non-orthogonal spin components:}\\
We computed the minimal uncertainty for a measurement of two spin-s components that span an angle $\phi$. Without loss of generality we can assume one of the components to be given by $L_z$ and the other one to lie in the $L_z-L_x$ plane. So we can take 
\begin{align}
L_\phi=\cos{\phi}L_z+\sin{\phi}L_x.
\end{align}
The value of the uncertainty bound in dependence of the angle $\phi$ is shown in Fig.~\ref{angularplot} and Tab.~\ref{angulartab}.
\quad
\quad
\begin{table}[h!]
\begin{tabular} {c|ccccc}
\text {spin/angle} & $0$ &$\frac {\pi} {8}$ &$\frac {\pi} {4} $&$\frac {3\pi} {8}$ &$\frac {\pi} {2} $ \\\hline\\
 1 & 0 & 0.0378 & 0.1431 & 0.2910 & 0.4365 \\ \\
 2 & 0 & 0.0743 & 0.2754 & 0.5318 & 0.7478 \\ \\
 3 & 0 & 0.1108 & 0.3984 & 0.7444 & 1.0131 \\ \\
\end{tabular}
\caption{Numerical values for the uncertainty of non-orthogonal spin components $L_z$ and $L_\phi$. Due to periodicity only angles from the interval $ [ 0,\pi/2  ] $ are relevant.\label{angulartab} }
\end{table}

{\bf Entanglement detection with local noise}\\

\begin{figure}[b]
\begin{minipage}[c]{0.9\linewidth}
\begin{flushleft}
\def\svgwidth{0.9\linewidth}
\begingroup%
  \makeatletter%
  \providecommand\color[2][]{%
    \errmessage{(Inkscape) Color is used for the text in Inkscape, but the package 'color.sty' is not loaded}%
    \renewcommand\color[2][]{}%
  }%
  \providecommand\transparent[1]{%
    \errmessage{(Inkscape) Transparency is used (non-zero) for the text in Inkscape, but the package 'transparent.sty' is not loaded}%
    \renewcommand\transparent[1]{}%
  }%
  \providecommand\rotatebox[2]{#2}%
  \ifx\svgwidth\undefined%
    \setlength{\unitlength}{539bp}%
    \ifx\svgscale\undefined%
      \relax%
    \else%
      \setlength{\unitlength}{\unitlength * \real{\svgscale}}%
    \fi%
  \else%
    \setlength{\unitlength}{\svgwidth}%
  \fi%
  \global\let\svgwidth\undefined%
  \global\let\svgscale\undefined%
  \makeatother%
  \begin{picture}(1,0.33209647)%
    \put(0,0){\includegraphics[width=\unitlength]{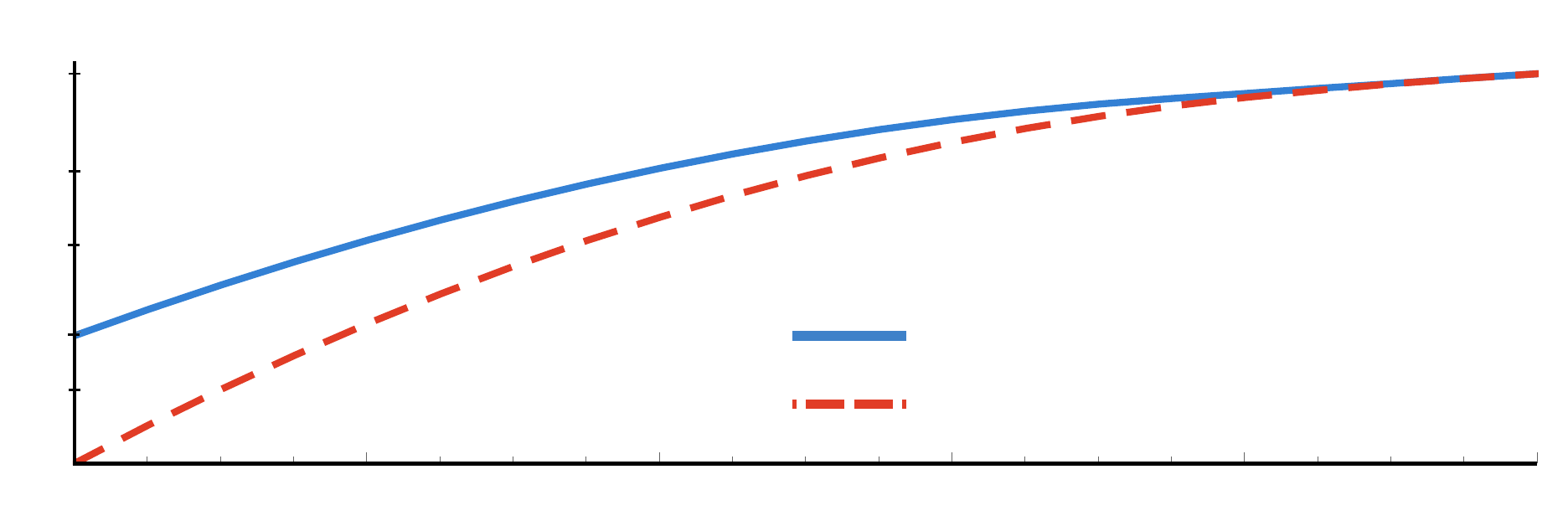}}%
    \put(-0.05,0.28074705){\color[rgb]{0,0,0}\makebox(0,0)[lb]{\smash{\small$8/3$}}}%
    \put(-0.05,0.11245181){\color[rgb]{0,0,0}\makebox(0,0)[lb]{\smash{\small$7/8$}}}%
    \put(0.01852656,0){\color[rgb]{0,0,0}\makebox(0,0)[lb]{\smash{\small$0$}}}%
    \put(0.21456693,-0.01){\color[rgb]{0,0,0}\makebox(0,0)[lb]{\smash{\small$0.2$}}}%
    \put(0.3972055,-0.01){\color[rgb]{0,0,0}\makebox(0,0)[lb]{\smash{\small$0.4$}}}%
    \put(0.58443393,-0.01){\color[rgb]{0,0,0}\makebox(0,0)[lb]{\smash{\small$0.6$}}}%
    \put(0.77166244,-0.01){\color[rgb]{0,0,0}\makebox(0,0)[lb]{\smash{\small$0.8$}}}%
    \put(0.96348071,-0.01){\color[rgb]{0,0,0}\makebox(0,0)[lb]{\smash{\small$1$}}}%
    \put(0.59369202,0.115){\color[rgb]{0,0,0}\makebox(0,0)[lb]{\smash{\small separable states}}}%
    \put(0.59369202,0.065){\color[rgb]{0,0,0}\makebox(0,0)[lb]{\smash{\small entangled states}}}%
    \put(0.5,0.17){\color[rgb]{0,0,0}\makebox(0,0)[lb]{\smash{
    {\small $\var{\rho}{M_1^\alpha} +\var{\rho}{M_2^\alpha}$}
    }}}%
    \put(1,0.03){\color[rgb]{0,0,0}\makebox(0,0)[lb]{\smash{$\alpha$}}}%
  \end{picture}%
\endgroup%
\end{flushleft}
\end{minipage}
\caption{\label{fig:lubed}
 Linear uncertainty bounds with equal weights in dependence of local noise, evaluated for measurements $M_1$ and $M_2$ on separable and entangled states.
Any state that yields a variance sum below the blue solid line is entangled.
}
\end{figure}
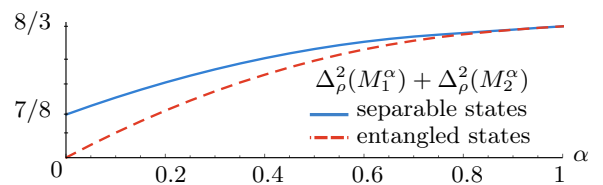

\end{document}